# Atomistic understanding of hydrogen coverage on $RuO_2$(110) surface under electrochemical conditions from *ab initio* statistical thermodynamics

Lei Zhang[1], Jan Kloppenburg[2], Chia-Yi Lin[3], Luka Mitrovic[4], Simon Gelin[5], Ismaila Dabo[5], Darrell G. Schlom[4], Jin Suntivich[4], Geoffroy Hautier[1,*]

[1] Thayer School of Engineering, Dartmouth College
[2] Department of Chemistry and Materials Science, Aalto University
[3] Smith School of Chemical and Biomolecular Engineering, Cornell University
[4] Department of Materials Science and Engineering, Cornell University
[5] Department of Materials Science and Engineering, The Pennsylvania State University

## Abstract

Understanding the dehydrogenation of transition metal oxide surfaces under electrochemical potential is critical to the control of important chemical processes such as the oxygen evolution reaction (OER). Using first principles computations, we model the thermodynamic dehydrogenation process on $RuO_2$(110) and compare the results to experimental cyclic voltammetry (CV) on single crystal. We use a cluster expansion model trained on *ab initio* energy data coupled with Monte Carlo (MC) sampling to derive the macroscopic electrochemical observables, i.e., experimental CV, from the energetics of different hydrogen coverage microstates on well-defined $RuO_2$(110). Our model reproduces the unique "two-peaks" cyclic voltammetry observed experimentally with current density peak positions and shapes in good qualitative agreement. We show that $RuO_2$ (110) starts as a water-covered surface with hydrogen on bridge (BRG) and coordination-unsaturated sites (CUS) at low potential (<0.4 V vs. reversible hydrogen electrode, RHE). As the potential increases, the hydrogens on BRG desorb, becoming the main contributor to the first CV peak with smaller contributions from CUS. When all BRG hydrogens are desorbed (before 1.2V vs RHE), the remaining CUS hydrogens desorb abruptly in a very small potential window leading to the sharp second peak observed during CV. Our work shows that above 1.23V, the OER proceeds on a fully dehydrogenated $RuO_2$(110) surface. We also demonstrate that the electrochemical dehydrogenation process on rutile involves multiple sites in a complex sequence of dehydrogenation. Our work highlights the use of first principles techniques coupled with statistical mechanics to model the electrochemical behavior of transition metal oxides surfaces.

## Introduction

Electrochemical water-splitting is a viable way of producing sustainable, green hydrogen[1,2]. However, catalysts that can efficiently drive the oxygen evolution reaction (OER), while maintaining reasonable stability are still under investigation and development[3]. OER catalysts that can sustain both acidic and anodic corrosions are very limited according to the Pourbaix diagrams[4,5]. Only oxides from the precious-metal group[6], e.g. RuO$_2$[7], IrO$_2$[8] and their derivatives[9–11], have met both the activity and stability requirements.

Aiming at improving the activity and stability of Ru and Ir-based catalysts, various catalyst-modifying strategies, e.g. doping or alloying[12], ligands engineering[13,14], nano-engineering[15,16], have been reported. However, the atomic-level reaction mechanism especially how the metal oxides exchange hydrogen with the solvent before and during the OER is still an open question. Recent experiments on well-defined crystals with clear surface terminations[17–20] using advanced thin film deposition, e.g. molecular beam epitaxy (MBE)[21] and pulsed laser deposition (PLD)[22,23] have provided insights into this fundamental question. Cyclic voltammetry (CV)[21] provides a direct probe of the surface reaction, which can be coupled to physical characterizations based on optical[24], or electronic[14] response or atomic force or scanning tunneling microscopy[17].

We have previously reported CV of rutile RuO$_2$(110) films grown using MBE[25,26]. Before the onset of the OER, two peaks were observed: one broad (0.8-1.0 V vs. reversible hydrogen electrode, "RHE") and another narrow (1.3 V vs. RHE). The CV before the OER is indicative of the current coming from the surface dehydrogenation. Starting from the fully hydrogenated RuO$_2$(110) surface, the hydrogen adsorbed species are removed as the voltage increases. The presence of the two peaks on RuO$_2$(110) is different from the one peak water dehydrogenation feature to form *OH on Pt single crystal CV[27]. The two peaks were also observed on a rutile IrO$_2$(110) film[20]. Using density functional theory (DFT) computations, the first peak was attributed to adsorbed water oxidation to hydroxide (*H$_2$O → *OH + H$^+$ + e$^-$) and the second peak to hydroxide dehydrogenation (*OH → *O + H$^+$ + e$^-$). However, this previous theoretical analysis of the CV only considered hydrogen adsorbate patterns in small supercells as it is commonly performed in the field[6]. Realistic surfaces under electrochemical potential should consider much larger supercell for the adsorbates and include entropic effects. A better model of the surface dehydrogenation would require considering the complexity of the surface including enumerating possible microstates (hydrogen coverage configurations) and statistically sampling them at different electrochemical potential. One step in this direction would be to assume a mean-field type of interaction with interaction parameters fitted from experiments[28] (e.g., Frumkin isotherm), or uses a simplified sampling using a partition function approach with limited configuration space[29]. A more elaborate technique to model the statistical mechanics of the different adsorbate configurations is the cluster expansion method[30] which comprehensively capture the adsorbate-adsorbate interactions using effective interaction parameters are fitted directly from DFT calculations. Cluster expansion models have been widely used to understand the phase transformation in structural alloys[31], Li-ion battery cathode materials[32] and bimetallic alloy catalysts[33]. However, its application to surface modeling under electrochemical potential has been limited with the notable exception of a simple cluster expansion (pair-wise interactions) used to model the CV of a Pt single crystal surfaces[27].

Here, we fit a cluster expansion model based on DFT data for different configurations of hydrogen on RuO$_2$(110) and sample the surface microstates using statistical mechanics coupled with a Monte Carlo (MC) algorithm at different potentials. Our model can qualitatively reproduce the two peaks in the experimental CV with the right peak positions and broadness. This agreement allows us to assign the CV peaks to specific surface dehydrogenation processes and clarify the atomistic mechanisms at play on $RuO_2$ (110) right before OER.

## Methods

The density functional theory (DFT) calculations are performed using VASP, a plane-wave basis code employing projector augmented wave (PAW) pseudopotentials with the implicit solvation model version, i.e. VASPsol[34,35]. Both dielectric and ionic responses upon the $RuO_2$(110) surface are considered by specifying the dielectric constant (78.4) and Debye length (9.613 Å) of the 0.1 M NaOH aqueous solution, where the experimental CV benchmark is based. Calculations are preformed using the generalized gradient approximation (GGA)-Perdew-Burke-Ernzheroff (PBE) functional and the PAW_PBE H (15Jun2001), PAW_PBE O (08Apr2002) and PAW_PBE Ru_pv (28Jan2005) pseudopotentials. A 4-Ru layer slab is constructed with an O-terminated BRG and CUS layer for *H adsorption. The bottom layer is with Ru termination to maintain the overall bulk $RuO_2$ stoichiometry. The vacuum size perpendicular to the surface is set as 10 Å. A larger vacuum space, e.g. 15 Å, shift the total energy less than 40 meV and does not affect the statistical thermodynamics. An "all band simultaneous update of orbitals" (ALGO=All) method is used to find the electronic ground state. The bottom two O-Ru-O layers are frozen while the top layers are allowed to relax during the slab calculations. A relaxed 1*1 slab with in-plane dimension of 2 Ru atoms per layer is used as prototype for the primitive cell. For primitive cells, a 6*3*1 Gamma-centered k-points was used. K-points for larger supercells are adjusted accordingly to maintain a constant k-point density. 400 eV energy cutoff are used together with an energy convergence criterion of $10^{-4}$ eV/unit-cell. Cluster expansion is performed using CASM[36]. Cluster expansion is an energy Hamiltonian depending only on lattice occupations. The formula is shown in Equation 1:

*Equation 1* $E(\boldsymbol{\sigma}) = \sum_\alpha m_\alpha J_\alpha \langle \prod_{i \in \alpha'} \sigma_i \rangle$

where $\sigma_i$ is a spin-like variable of up or down (occupation variable of -1 or +1 for an occupied or unoccupied atom specie) and is an element of vector $\sigma$ representing configuration space; Coefficient $J_\alpha$ is a fitted energy value for cluster $\alpha$ and called effective cluster interaction (ECI); $m_\alpha$ is the multiplicity of symmetrically equivalent cluster $\alpha$, the group of equivalent cluster $\alpha$ is denoted as $\alpha'$. It can couple with MC sampling to evaluate thermodynamic quantities and predict phase transformations.

To train the cluster expansion model, over 500 supercells up to eight-times of the primitive cell prototype with various lattice vectors and *H adsorption occupations are enumerated and calculated by DFT. The dataset is then used to train a cluster-based expansion model of total energy using selected interactions within a series of doublet, triplet and quadruplet clusters with diameters of 9, 7 and 6 Å. A Metropolis MC under the fixed chemical potential of hydrogen species (grand-canonical ensemble) is then performed using the trained cluster-expansion model to evaluate the free energy of a 64×64 supercell, yielding statistically robust thermodynamic quantities by taking into account both enthalpy and configuration entropy of microstates, serving a computational counterpart of experimental CV. Chemical potentials of hydrogen are scanned, which directly couples with the applied potential $U$ using the computational hydrogen electrode (CHE) model[37]. The way of using $G_{H_2}$ to represent the energy of a coupled proton and electron under the reversible condition (298.15 K, 1 atm, arbitrary pH) is well-established by J. Norskov et. al. in the CHE. Thus, any other electrode potentials can be referenced back to the CHE's absolute potential, i.e., 4.44 V:

$$Equation\ 2\ U(w.r.t\ RHE) = 4.44 - 0.5 * G_{H_2} - \mu_H$$

where 4.44 is the absolute electrode potential of RHE, $G_{H_2}$ is the Gibbs free energy of a H$_2$ molecule including 0 K DFT energy (calibrated to the NIST JANAF table), zero-point vibrational energy, and free energy from 0 to 298.15 K with data from the NIST JANAF table[38]; $\mu_H$ is the chemical potential of H within the chemical space of Ru$_8$O$_{16}$–Ru$_8$O$_{16}$H$_3$ pseudo-binary governed by the thermodynamics of *H on RuO$_2$ (110). It is worth mentioning that the vibrational (including the zero-point energy, i.e. ZPE) contribution to the free energy of *H adsorption can be safely neglected as it is configuration and coverage independent, which cancels out itself when evaluating the free energy of mixing within the convex hull. Therefore, the system's mixing thermodynamics is governed mainly by 0 K enthalpies and configuration entropies of relevant microstates.

The electrochemical experiments were conducted with 47 formula-units thick molecular beam epitaxy (MBE)-grown RuO$_2$(110) films (~ 19 nm) in 0.1 M NaOH. The MBE synthesis was conducted in Veeco Gen 10 MBE. The ruthenium flux was obtained by evaporating Ru metal (99.99 % pure, ESPI Metals) using an electron beam into an ozone environment to form RuO$_2$ on TiO$_2$(110) substrates at 350 °C. Ozone background pressure was maintained at 10$^{-6}$ Torr during the heating up of substrates, growth, and cooling of the sample. The MBE-grown films were confirmed by X-ray diffraction, as shown in Figure S 1.

The MBE-grown RuO$_2$(110) films were then prepared into a working electrode by attaching Titanium wire on RuO2 films using Gallium liquid metal (99.99% pure, Amazon), and silver paint (Ted Pella, Leitsilber 200) and dried in air overnight. After that, the epoxy (Omegabond 101) was used to cover the silver paint, edges, and the backside of sample except for the RuO$_2$(110) film. The RuO$_2$(110) electrode was rinsed in DI water right before electrochemical experiments. A Pt wire was used as counter electrode. The reference electrode was reverse hydrogen electrode (RHE). The 0.1 M NaOH electrolyte was prepared by dissolving NaOH (99.99 % Suprapur® from Millipore Sigma) in ultrapure water (18.2 MΩ-cm). The CV was obtained in an Ar-saturated electrolyte with scan rate at 50 mV/s.

## Results

**Experimental cyclic voltammetry and relation to hydrogen coverage**

Figure 1 shows the experimental CV data on RuO$_2$(110) single crystal thin film. The measurement is performed at a potential scan rate of 50 mV/s. Current density is then measured through the potential with the sample submerged in a 0.1M NaOH aqueous electrolyte (pH = 13). Two peaks are clearly identified with a broad 1st peak spanning from 0.4 to 1.2 V and a sharp 2nd peak centered around 1.3V vs. RHE. No hysteresis is found in the cyclic measurement, indicating a spontaneous and reversible hydrogen desorption/adsorption behind these two peaks, consistent with our previous measurements on Ru and Ir-based rutile (110) surface[20,21]. We can interpret the CV data as a measure of the current related to dehydrogenation of the oxide surface. Mathematically, the current density $dQ/dt$ and voltage scan rate $dU/dt$ are connected by Equation 3 assuming a satisfied equilibrium condition:

$$\text{Equation 3} \quad \frac{dU}{dt} = \frac{dU}{d\theta}\frac{d\theta}{dQ}\frac{dQ}{dt} \text{ assuming } \Delta G(U, \theta) = 0$$

where $\theta$ is the *H coverage. The transferred charge $Q$ is linearly proportional to $\theta$ as each hydrogen removed from the surface will release an electron. By integrating current density $i$ w.r.t the applied potential $U$, equilibrium surface *H coverage is obtained as a function of $U$:

$$\text{Equation 4} \quad \theta = \int \frac{1}{KQ_{tot}} i \, dU$$

where $K$ is the potential scan rate and $Q_{tot}$ is the total charge transferred within the whole dehydrogenation process.

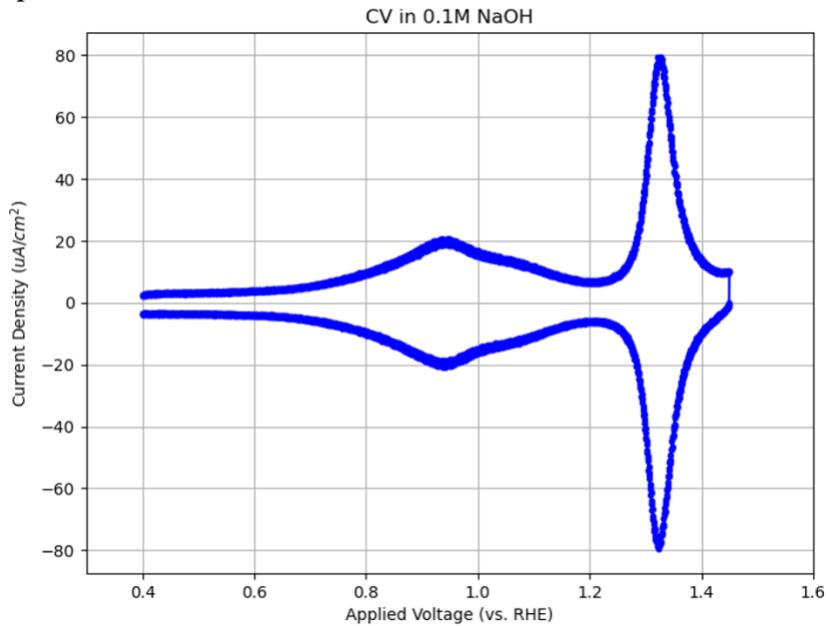

*Figure 1. Experimental cyclic voltammogram on single crystal thin film RuO$_2$(110) with measured current density at a potential scan rate of 50mV/s under 0.1M NaOH with surface normalization. Feature of two peaks: 1st broad peak located below 1.2 V with a width of ~1 V, 2nd sharp peak at ~1.3 V.*

**Cluster expansion**

First principles computations can be used to identify the atomistic mechanisms responsible for the change in hydrogen coverage and thus the CV. In previous work, we have performed such analysis computing with DFT a series of simple ordered configuration for surface hydrogens within a small unit cell[20,21]. This simple analysis attributed the first peak to the *$H_2O$ → *OH + $H^+$ + $e^-$ reaction and the second peak to the *OH → *O + $H^+$ + $e^-$ reaction. However, this approach is severely limited by the unit cell size and number of hydrogen coverage configurations where more complex hydrogen interactions with longer distances cannot be captured. Moreover, this previous work focused on the CUS sites which is known to be the active site for OER but the BRG site could also be part of the dehydrogenation process. A more realistic model requires a much larger system modeled by a supercell to reach the thermodynamic limit and a comprehensive MC sampling of many different hydrogen configurations over multiple adsorption sites. These larger systems are typically not within reach of DFT because of computational cost. However, the cluster expansion technique can be used to fit a model Hamiltonian that will provide an energy for any configuration of interest. When this cluster expansion model is coupled with MC sampling, microstates can be efficiently sampled and experimental observables (e.g., hydrogen coverage versus potential) can be obtained by theory.

Our model $RuO_2$ (110) surface has two types of O sites that can potentially bind with H, i.e. CUS and BRG, where in particular O on CUS can bind with 2 H to form a water molecule-like motif and BRG O can only accommodate 1 H (left of Figure 2). One primitive cell of $RuO_2$ (110) has 2 CUS and 1 BRG sites hosting 3 *H in total. The O-H bond on BRG can point to the lattice vector direction of **b** or **-b**. The two CUS O-H bonds have CUS1 pointing along with the BRG O-H and CUS2 O-H slanted towards lattice vector **a** or **-a**. To avoid the strong steric repulsion among *H, 1 ML (i.e. monolayer) *H have a rotationally rigid pattern with a restricted rotational degree of freedom. Different rotationally ordered patterns were tested and found energetically degenerate, hence are not considered explicitly in the statistical sampling. The way cluster expansion model works is to first enumerate a sufficient size of training set with various configurations based on the primitive slab. Figure 2 shows a few representative configurations of hydrogen in a 2×2 supercell. Lattice vectors and configurations are allowed to vary during the fit of the cluster expansion to ensure a diverse sampling and in total around 500 structures of different supercell sizes have been generated and computed with DFT. For these 500 structures, we use DFT with an implicit solvation model to compute energies (see Methods). This data set is then used to train a cluster expansion model that links the configuration of hydrogens (i.e., what sites are occupied or not, through an occupation variable $\sigma$ that is -1 or 1, elements of occupation matrix **$\sigma$** with lattice sites and clusters as row and column indices) and their corresponding energies (right of Figure 2). Quantities learned through this process are the effective cluster interactions $J_\alpha$ for a series of clusters (i.e., groups of sites) that can be doublet, triplet, quadruplet etc… Mathematically, it is an inverse matrix problem where the interaction parameters $J_\alpha$ are solved numerically under the given the energy vector $E(\sigma)$ as a function of the occupation matrix **$\sigma$**. Therefore, rank of matrix **$\sigma$** needs to be as close to full as possible and the symmetrically distinct cluster basis sets (called orbit $\alpha$) needs to be sufficient to describe the interactions. A robust cluster expansion model captures the physics of the system's interactions by using a succinct set of interaction parameters with expanded clusters (i.e. doublet, triplet, quadruplet, etc.) while avoiding overfitting.

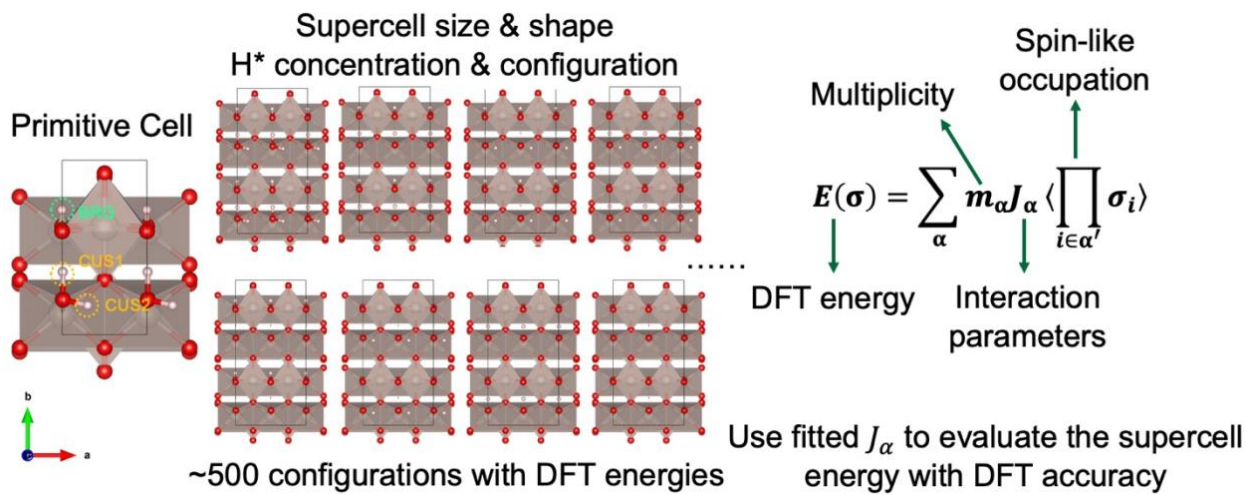

*Figure 2. A schematic showing the random enumeration of DFT training sets (only 2\*2 supercells with a few representative \*H occupation states are shown for demonstration, while various sizes and shapes of supercells with \*H coverages and configurations were enumerated) using the primitive cell on the left as the prototype. A vector-matrix projection is then constructed between the supercell energy vector $E(\sigma)$ and the occupation matrix $\sigma$ (with cluster basis set and lattice site indices). Interaction parameters $J_\alpha$ within a symmetrically distinct cluster group (or called as orbit) α are then fitted using the energy and occupation as input. A well-trained set of $J_\alpha$ constitutes a so-called cluster expansion model that is ready for efficient free energy sampling of a much larger supercell that is not accessible through quantum mechanics calculations like DFT.*

**Cluster Expansion and comparison to DFT Convex Hull**

The zero K DFT mixing energies (blue circles) of the configurations studied in this work are plotted versus the hydrogen fraction in Figure 3. Here, a convex hull is constructed by connecting ground states (blue dots) within the chemical space. DFT predicts three important ground states in the middle with *H site fraction of 0.833 (5/6 ML), 0.666 (4/6 ML) and 0.5 (3/6 ML). The formation energy is over -0.4 eV/primitive cell, a significantly larger value than $k_B T$ at room temperature (0.0259 eV), indicating a strong tendency of *H mixing over the surface. However, between 0.5 ML and 0 ML, no *H configuration formed more stable ground states than the convex line, indicating a possible sharp phase transition during dehydrogenation in this *H coverage range.

The formation energy from a trained cluster expansion is also plotted (red dots) showing a good agreement with the data from DFT (blue circles). A combination of doublet, triplet and quadruplet clusters interactions within diameters of 9, 7 and 6 Å were found to give the best fitting quality, with root mean squared error (rmse) and cross-validation scores less than 2 meV/atom. More importantly, the overall shape of the convex hull and the key ground states are well-reproduced. Since microstates are sampled only at room temperature, data with formation energies close to the hull are most relevant in the following MC sampling.

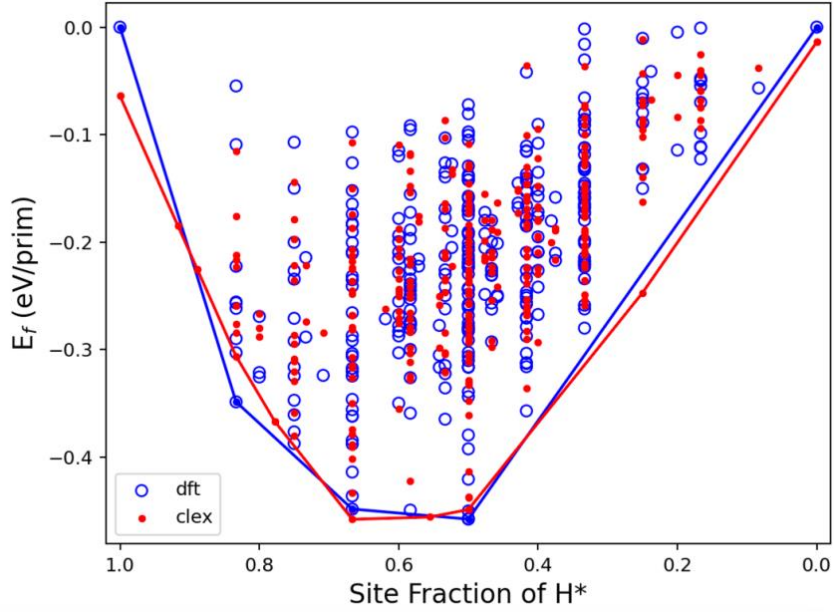

*Figure 3. The 0 K formation energy in eV per primitive cell (blue circle: DFT calculations, red dot: cluster expansion prediction). The ground state with a lowest formation energy of ~-0.47 eV is on 2/3 ML (0.666) coverage *H. The rest of deep ground states locate at 5/6 and 1/2 ML *H coverages. The root mean squared error (rmse) and cross-validation (cv) scores are both less than 2 meV/atom. The clex model is trained using a 9-7-6 Å interaction cutoff lengths with a doublet-triplet-quadruplet cluster combination.*

**Dehydrogenation profile from Monte-Carlo Simulations**

The convex hull presented in Figure 3 is a zero Kelvin picture of the dehydrogenation process on $RuO_2$(110) surface. Introducing temperature effects and entropy requires to move to statistical mechanics and sample the equilibrium average coverage at different hydrogen chemical potentials (i.e., different electrochemical potentials in an electrochemical settings). This average requires the energy of many different configurations of hydrogen to be evaluated in large supercells, something impossible to achieve directly from DFT. However, we can use our cluster expansion fitted on DFT data to evaluate any configuration of hydrogen and use the laws of statistical mechanics.

The equilibrium *H coverage with the detailed *H site-fractions are computed w.r.t to the hydrogen chemical potential ($\mu_H$). The ensemble average, i.e., the system's grand potential at a given $\mu_H$ and temperature $T$ is evaluated through the partition function $Z(\mu_H, T)$ defined in *Equation 5*:

$$\text{Equation 5 } Z(\mu_H, T) = \sum_i Z_i = \sum_i exp\left(\frac{N_{*H,i}\mu_H - E_i}{k_B T}\right)$$

where $N_{*H,i}$ and $E_i$ are the number of *H and total energy of the microstate $i$. Probability $P_i$ of microstate $i$ is given as $Z_i/Z$ with $Z_i$ the $i$-th component of $Z$.

Grand potential $\Phi_G$ is related with the partition function by *Equation 6*:

$$\text{Equation 6 } -k_B T \ln Z = \Phi_G = \langle E \rangle - TS - \mu_H \langle N_{*H} \rangle$$

where total energy $\langle E \rangle$, number of *H $\langle N_{*H} \rangle$ and entropy $S$ are averaged values by sampling sufficient microstates:

$$\text{Equation 7} \quad \langle E \rangle = \sum_i E_i P_i = \frac{1}{Z}\sum_i E_i \, exp\left(\frac{N_{*H,i}\mu_H - E_i}{k_B T}\right)$$

$$\text{Equation 8} \quad \langle N_{*H} \rangle = \sum_i N_{*H,i} P_i = \frac{1}{Z}\sum_i N_{*H,i} \, exp\left(\frac{N_{*H,i}\mu_H - E_i}{k_B T}\right)$$

$$\text{Equation 9} \quad S = -k_B P_i \ln P_i$$

In practice, we have computed these quantities using our cluster expansion (E(σ)) and Metropolis MC importance sampling on a 64×64 $RuO_2$(110) surface containing multiple hydrogen sites. Our model is not limited by small unit cell size and includes the configurational entropy of hydrogen configuration. The MC sampling provides the coverage vs hydrogen chemical potential and thus electrochemical potentials as they are linked by the hydrogen reference electrode $H^+ + e^- \rightarrow H_2$ (see Methods).

To facilitate the comparison between theory and experiment, we have integrated the current from the experimental CV to obtain the experimental hydrogen coverage vs. potential (blue curve) and compare it to the results of our MC simulations (red curve) in Figure 4. The 1$^{st}$ stage of the dehydrogenation process occurs in our model between 0.3~1.2V (~1 V potential window) where 50% of hydrogen is desorbed. This 1$^{st}$ stage of dehydrogenation corresponds to the 1$^{st}$ broad peak in the experimental CV (see Figure 1 as well). The remaining 50% hydrogen gets desorbed from ~1.2 to 1.3V within a small 0.1 V potential window corresponding to a much sharper 2$^{nd}$ peak (see Figure 1 as well). The agreement between theory and experiment is good overall.

Although the experiment shows a smooth transition during the 1$^{st}$ peak, the model shows some staging. We attribute this discrepancy to the limitation of the implicit solvation model that neglects the entropic contribution of the explicit solvent water molecules within the Helmholtz layer. We expect the explicit solvent molecules to promote interfacial disorders and smear the simulated steps within the 1$^{st}$ peak. Nonetheless, the "two-peak" feature in the experimental CV with the 1$^{st}$ one broader and 2$^{nd}$ one sharper is very well reproduced in our model. The specific potential windows for the two peaks also agree reasonably well with the experiment. The peaks show less than 200 mV shift between simulation and experiment. This mismatch could be due to the used PBE functional. As we have previously discussed, errors up to 100 meV are not uncommon for oxidation processes in transition metal oxides within semilocal functionals such as GGA[21]. The neglect of the interfacial electric field could also explain the discrepancy. Moreover, rotational degrees of freedom (rotational configuration entropy) are higher for surfaces with lower *H coverage, as they have larger space necessary for OH rotations without steric hindrance from nearby hydroxyl groups. Water molecules near an inorganic surface are expected to be more disordered, rendering an entropically destabilized state at low *H coverage. Those missing factors can shift the dehydrogenation profile (together with the peaks) towards a higher applied potential.

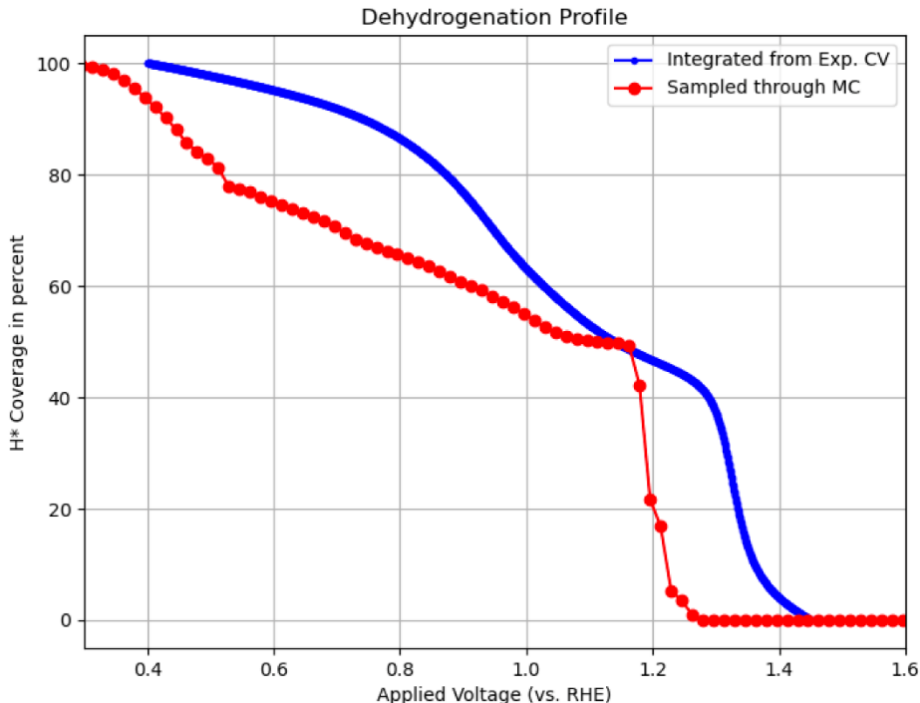

*Figure 4. Dehydrogenation profile on RuO$_2$ (110) under the applied voltage (vs. a reversible hydrogen electrode) integrated from the experimental cyclic voltammetry (blue curve) and sampled from the Monte-Carlo simulation (red dotted line). A slow dehydrogenation process is revealed below ~1.2 V, after which the rest ½ ML *H undergoes a sharp desorption from ~1.2 to 1.4 V. All *H are desorbed further beyond, rendering a *H clean surface before the onset of OER.*

To further understand the atomistic details behind the dehydrogenation, we isolate the site-specific *H on BRG and CUS shown in Figure 5 (see also Figure S 3). As the potential increases from 0.3 V to 0.5 V vs. RHE in the very beginning, *H on CUS (orange curve) get dehydrogenated first by 1/6 ML *H. However, after this point and until the end of the 1$^{st}$ stage, no further *H on CUS participates in additional dehydrogenation. Instead, BRG *H (green curve) contributes more to the dehydrogenation. This contribution from BRG *H continues to the end of 1$^{st}$ stage, where all BRG *H is desorbed, contributing to 1/3 ML of overall *H coverage. The 2$^{nd}$ stage starts at a potential of ~1.2 V vs. RHE. Here, the rest ½ ML *H on CUS is desorbed in a rather narrow potential window in stark contrast to the 1$^{st}$ dehydrogenation stage.

Our finding suggests that the "two-stage" dehydrogenation occurs via a "solid-solution-like" behavior among *H on BRG during the 1$^{st}$ dehydrogenation stage and a "two-phase-like" behavior of *H on CUS at ~1.2V between ½ and 0 ML coverage phase regions. Interestingly, the sharp desorption on BRG and CUS at 0.5 V might be due to the large slope of the convex hull near the full *H coverage. This complex hull behavior from DFT and the cluster expansion model could give this small desorption an appearance of a "kink." However, this does not alter our conclusion about the "two-peak" feature. *H redistribution between BRG and CUS is observed at 0.5 V. The suspension of desorption on CUS after 0.5 V with ½ ML *H is due to a similar *H redistribution between CUS1 and CUS2 (See Figure S 3). This counter-intuitive *H redistribution was reported before, e.g. *H could hop from site to site facilitated by interfacial water layers through the Grotthuss mechanism[39,40]. This detailed site-fraction of *H unveils the possibility of adsorbate redistribution during a trivial desorption process where a monotonic *H coverage evolution is concerned.

We confirm that, before the onset of OER (above 1.3 or 1.4 V), there is no remaining *H on the surface, indicating a decoupling of the first two steps (*OH$_2$ → *OH → *O) with the last two steps (*O → *OOH → *OO) in the 4-step 4e$^-$-involved OER. This finding might be of importance for modeling the kinetics of the OER process as it shows that the *O + H$_2$O → *OOH + H$^+$ + e$^-$ reaction occurs on an oxygen terminated, free of hydrogen (110) surface in RuO$_2$. Therefore, a simple *OH$_2$ → *OH → *O reaction scheme should be revised to a more detailed one, i.e. *OH$_{2CUS}$*OH$_{BRG}$-*OH$_{2CUS}$*OH$_{BRG}$ → *OH$_{2CUS}$*OH$_{BRG}$-*OH$_{CUS}$O$_{CUS}$*OH$_{BRG}$ → *OH$_{2CUS}$*O$_{BRG}$-*OH$_{CUS}$O$_{CUS}$O$_{BRG}$ → O$_{2CUS}$O$_{BRG}$-O$_{2CUS}$O$_{BRG}$, which will be further discussed next. Moreover, we emphasize that for non-trivial oxide surfaces, e.g., RuO$_2$ (110), complex *H interactions among different adsorption sites greatly complicate the picture of a single active site on which intermediates for the OER (OH, O, OOH) are computed.

Our work shows that the BRG site plays an important role in the CV and will influence the peaks positions. This is of importance as the 4-steps 4e$^-$ mechanism has linked catalyst activity to hydrogen adsorption energetics. The scaling relationship especially links activity to any of the free energy change in the catalytic steps (e.g., *OH to *O or *OH$_2$ to *OH). This naturally leads to using the CV peak positions that are related to hydrogen adsorption energetics as descriptors of catalytic activity[20,21]. However, the BRG site is often considered inactive in the OER catalytic cycle and ignored in the typical 4 steps 4e$^-$ mechanism. This makes a direct link between hydrogen coverage measurements through CV (driven by CUS and BRG) and the energy change of the 4-step which is typically driven by CUS only less trivial than previously thought.

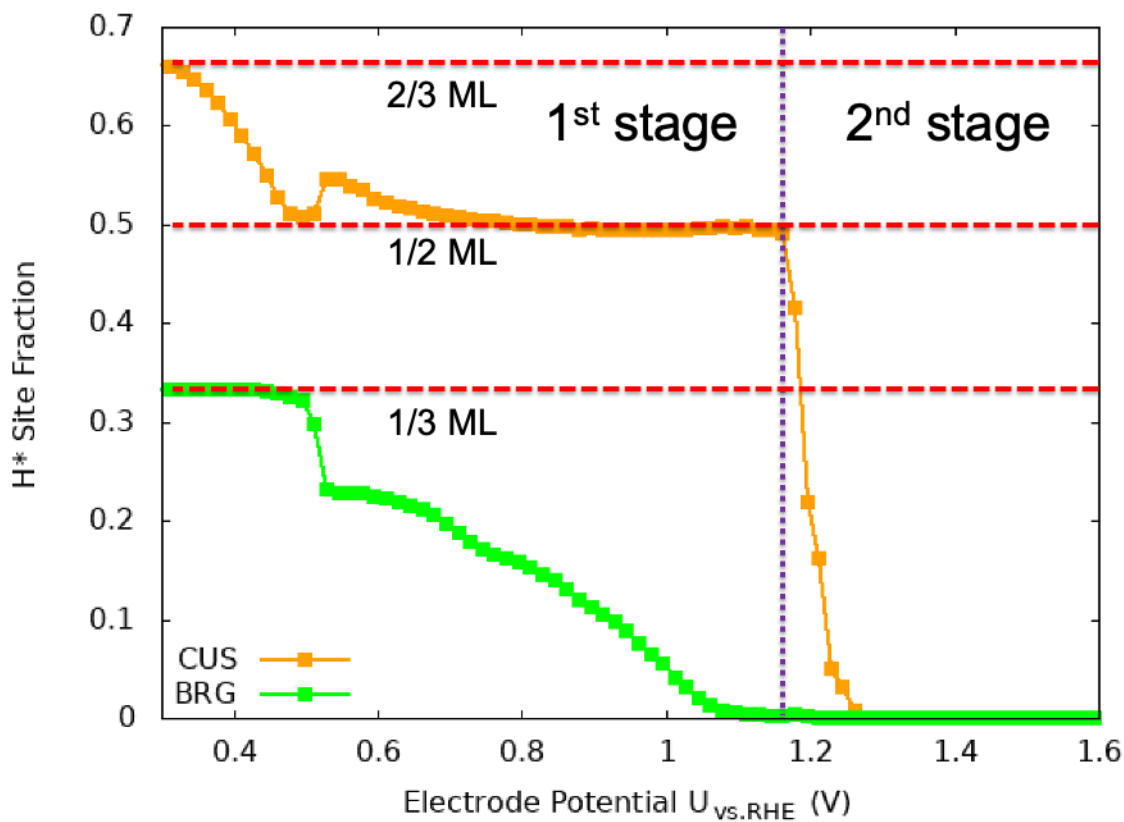

*Figure 5. Site-specific *H desorption in two stages. *H desorption profile over CUS and BRG, shown as orange and green ■. During the 1st peak voltage window, only 1/6 ML *H from CUS but all BRG *H desorb. CUS *H desorb first but suspend the process afterwards, leaving an unchanged ½ ML *H coverage. All BRG *H contributing 1/3 ML surface desorb gradually over the course of the 1st voltage window. During the 2nd peak voltage window, the remained ½ ML *H on CUS desorb quickly above a 1.2 V potential threshold. A slight *H redistribution is observed at ~0.5 V between CUS and BRG.*

To illustrate the evolution of *H configuration patterns, snapshots from MC during the potential scan is presented at representative *H coverages in Figure 6. It is found that *H on CUS forms a **2a*1b** superlattice with alternating *H and vacant * along **a** direction at 0.50 V with 5/6 ML coverage. At ~0.55 V, BRG *H starts to form a **3a*1b** superlattice where 1 out of 3 BRG *H is missing in each BRG row. Some *H get re-adsorbed on CUS, forming a similar **3a*1b** pattern. Higher applied potentials gradually strip off the rest of BRG *H while CUS *H remains a **2a*1b** superlattice of ½ ML coverage till 1.15V. Starting at 1.2 V the left CUS *H strips start to peel off from the surface "stripe-by-stripe" rather than "atom-by-atom", leaving fragmented CUS *H stripes with **2a** periodicity. This happens within a narrow potential window of 0.1 V. At and above 1.25 V all CUS *H are gone, leaving a clean surface with O-termination ready for another round of adsorption. In fact, unlike $TiO_2$, $RuO_2$ or $IrO_2$ have spontaneous dehydrogenation where overpotential is not needed in the first two OER steps[41]. We note that although *OOH is not included in the computational CV model, the CV is still well reproduced. This suggests that *OOH is likely not participating in the charge transfer process during the CV measurement below 1.23 V in our sample. This is different than previous results on PLD samples[42].

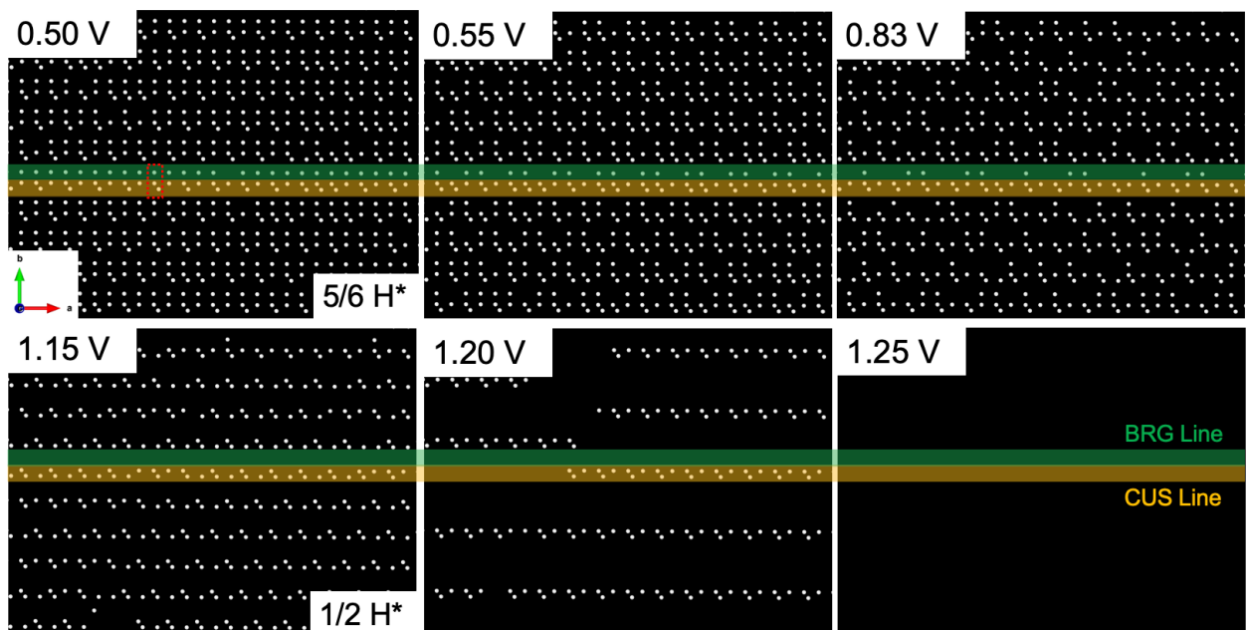

*Figure 6. Representative snapshots of the Metropolis Monte-Carlo simulation w.r.t applied potentials (vs. RHE) starting with the ordered patterns of *H: *$OH_{2CUS}$-*$OH_{BRG}$-*$OH_{2CUS}$-*$OH_{BRG}$. 5/6 ML *H appears in *$OH_{2CUS}$-*$OH_{BRG}$-*$OH_{CUS}$-*$OH_{BRG}$ at 0.50V where CUS *H are starting to desorb. With increasing potentials, BRG *H start to desorb as well and eventually are all gone at 1.15 V ($OH_{2CUS}$-$OH_{CUS}$). This belongs to the 1st stage of dehydrogenation with the potential window of almost 1 V. After that, all the rest CUS *H with ½ ML coverage desorb quickly within 0.1 V (1.15 to 1.25 V), by stripping off CUS *H stripe-by-stripe rather than atom-by-atom. *H on BRG and CUS form green and orange lines in the 64×64 MC supercell, with a dotted square encircling a 3 *H (2 on CUS and 1 on BRG) primitive cell (1st snapshot at 0.50 V).*

## Conclusion

We use a cluster expansion trained on DFT energies and a statistical mechanics approach to compute the CV and hydrogen coverage versus electrochemical potential for $RuO_2$(110). The model reproduces the two peaks observed in the experimental CV and provides insight into the atomistic mechanism that governs the shape and position of the two peaks. The 1st broad peak is due to the gradual dehydrogenation on the BRG sites and the partial (1/6 ML) dehydrogenation on CUS. The 2nd sharp peak is due to quick desorption of ½ ML *H from CUS and happens when all BRG sites have been dehydrogenated. Our work shows that at an electrochemical potential relevant to the OER (>1.2V vs RHE), the $RuO_2$(110) surface is fully dehydrogenated which can have important implications when modeling the kinetics of OER on realistic surfaces.

The application of a statistical thermodynamic model to connect the surface microscopic states with macroscopic experimental data allows a direct investigation of the atomistic dehydrogenation process on $RuO_2$(110). We conclude that this surface process follows an intricate process involving both sites on the rutile surface and complex interactions. While our model captures already much of the dehydrogenation behavior for $RuO_2$(110), future work could include other effects such as explicit solvent or electric field at the interface.

## Acknowledgement

This work was supported as part of the Center for Electrochemical Dynamics and Reactions at Surfaces (CEDARS), an Energy Frontier Research Center funded by the US Department of Energy, Office of Science, Basic Energy Sciences under award DE-SC0023415. The computational work used resources of the National Energy Research Scientific Computing Center (NERSC), a Department of Energy Office of Science User Facility using NERSC award BES-ERCAP0028800 and Andes8 clusters at Dartmouth College. The authors want to thank Dr. Wei Chen and Andrew Pike for helpful discussions.

*geoffroy.hautier@dartmouth.edu

## References

(1) Guan, D.; Wang, B.; Zhang, J.; Shi, R.; Jiao, K.; Li, L.; Wang, Y.; Xie, B.; Zhang, Q.; Yu, J.; Zhu, Y.; Shao, Z.; Ni, M. Hydrogen Society: From Present to Future. *Energy Environ. Sci.* **2023**, *16* (11), 4926–4943. https://doi.org/10.1039/D3EE02695G.

(2) Stamenkovic, V. R.; Strmcnik, D.; Lopes, P. P.; Markovic, N. M. Energy and Fuels from Electrochemical Interfaces. *Nat. Mater.* **2017**, *16* (1), 57–69. https://doi.org/10.1038/nmat4738.

(3) Kibsgaard, J.; Chorkendorff, I. Considerations for the Scaling-up of Water Splitting Catalysts. *Nat. Energy* **2019**, *4* (6), 430–433. https://doi.org/10.1038/s41560-019-0407-1.

(4) Zagalskaya, A.; Chaudhary, P.; Alexandrov, V. Corrosion of Electrochemical Energy Materials: Stability Analyses Beyond Pourbaix Diagrams. *J. Phys. Chem. C* **2023**, *127* (30), 14587–14598. https://doi.org/10.1021/acs.jpcc.3c01727.

(5) Wang, Z.; Guo, X.; Montoya, J.; Nørskov, J. K. Predicting Aqueous Stability of Solid with Computed Pourbaix Diagram Using SCAN Functional. *Npj Comput. Mater.* **2020**, *6* (1), 160. https://doi.org/10.1038/s41524-020-00430-3.

(6) Danilovic, N.; Subbaraman, R.; Chang, K.-C.; Chang, S. H.; Kang, Y. J.; Snyder, J.; Paulikas, A. P.; Strmcnik, D.; Kim, Y.-T.; Myers, D.; Stamenkovic, V. R.; Markovic, N. M. Activity–Stability Trends for the Oxygen Evolution Reaction on Monometallic Oxides in Acidic Environments. *J. Phys. Chem. Lett.* **2014**, *5* (14), 2474–2478. https://doi.org/10.1021/jz501061n.

(7) Nguyen, M.-T.; Mu, R.; Cantu, D. C.; Lyubinetsky, I.; Glezakou, V.-A.; Dohnálek, Z.; Rousseau, R. Dynamics, Stability, and Adsorption States of Water on Oxidized $RuO_2(110)$.

(8) Ha, M.-A.; Larsen, R. E. Multiple Reaction Pathways for the Oxygen Evolution Reaction May Contribute to $IrO_2$ (110)'s High Activity. *J. Electrochem. Soc.* **2021**, *168* (2), 024506. https://doi.org/10.1149/1945-7111/abdeea.

(9) Akbashev, A. R.; Roddatis, V.; Baeumer, C.; Liu, T.; Mefford, J. T.; Chueh, W. C. Probing the Stability of $SrIrO_3$ during Active Water Electrolysis *via Operando* Atomic Force Microscopy. *Energy Environ. Sci.* **2023**, *16* (2), 513–522. https://doi.org/10.1039/D2EE03704A.

(10) Wan, G.; Freeland, J. W.; Kloppenburg, J.; Petretto, G.; Nelson, J. N.; Kuo, D.-Y.; Sun, C.-J.; Wen, J.; Diulus, J. T.; Herman, G. S.; Dong, Y.; Kou, R.; Sun, J.; Chen, S.; Shen, K. M.; Schlom, D. G.; Rignanese, G.-M.; Hautier, G.; Fong, D. D.; Feng, Z.; Zhou, H.;

Suntivich, J. Amorphization Mechanism of SrIrO$_3$ Electrocatalyst: How Oxygen Redox Initiates Ionic Diffusion and Structural Reorganization. *Sci. Adv.* **2021**, *7* (2), eabc7323. https://doi.org/10.1126/sciadv.abc7323.

(11) Ben-Naim, M.; Liu, Y.; Stevens, M. B.; Lee, K.; Wette, M. R.; Boubnov, A.; Trofimov, A. A.; Ievlev, A. V.; Belianinov, A.; Davis, R. C.; Clemens, B. M.; Bare, S. R.; Hikita, Y.; Hwang, H. Y.; Higgins, D. C.; Sinclair, R.; Jaramillo, T. F. Understanding Degradation Mechanisms in SrIrO$_3$ Oxygen Evolution Electrocatalysts: Chemical and Structural Microscopy at the Nanoscale. *Adv. Funct. Mater.* **2021**, *31* (34), 2101542. https://doi.org/10.1002/adfm.202101542.

(12) Choubisa, H.; Abed, J.; Mendoza, D.; Matsumura, H.; Sugimura, M.; Yao, Z.; Wang, Z.; Sutherland, B. R.; Aspuru-Guzik, A.; Sargent, E. H. Accelerated Chemical Space Search Using a Quantum-Inspired Cluster Expansion Approach. *Matter* **2023**, *6* (2), 605–625. https://doi.org/10.1016/j.matt.2022.11.031.

(13) Hao, S.; Liu, M.; Pan, J.; Liu, X.; Tan, X.; Xu, N.; He, Y.; Lei, L.; Zhang, X. Dopants Fixation of Ruthenium for Boosting Acidic Oxygen Evolution Stability and Activity. *Nat. Commun.* **2020**, *11* (1), 5368. https://doi.org/10.1038/s41467-020-19212-y.

(14) Lee, S.; Lee, Y.-J.; Lee, G.; Soon, A. Activated Chemical Bonds in Nanoporous and Amorphous Iridium Oxides Favor Low Overpotential for Oxygen Evolution Reaction. *Nat. Commun.* **2022**, *13* (1), 3171. https://doi.org/10.1038/s41467-022-30838-y.

(15) Du, K.; Zhang, L.; Shan, J.; Guo, J.; Mao, J.; Yang, C.-C.; Wang, C.-H.; Hu, Z.; Ling, T. Interface Engineering Breaks Both Stability and Activity Limits of RuO2 for Sustainable Water Oxidation. *Nat. Commun.* **2022**, *13* (1), 5448. https://doi.org/10.1038/s41467-022-33150-x.

(16) Ping, X.; Liu, Y.; Zheng, L.; Song, Y.; Guo, L.; Chen, S.; Wei, Z. Locking the Lattice Oxygen in RuO2 to Stabilize Highly Active Ru Sites in Acidic Water Oxidation. *Nat. Commun.* **2024**, *15* (1), 2501. https://doi.org/10.1038/s41467-024-46815-6.

(17) Chang, S. H.; Danilovic, N.; Chang, K.-C.; Subbaraman, R.; Paulikas, A. P.; Fong, D. D.; Highland, M. J.; Baldo, P. M.; Stamenkovic, V. R.; Freeland, J. W.; Eastman, J. A.; Markovic, N. M. Functional Links between Stability and Reactivity of Strontium Ruthenate Single Crystals during Oxygen Evolution. *Nat. Commun.* **2014**, *5* (1), 4191. https://doi.org/10.1038/ncomms5191.

(18) Over, H. Fundamental Studies of Planar Single-Crystalline Oxide Model Electrodes (RuO$_2$, IrO$_2$) for Acidic Water Splitting. *ACS Catal.* **2021**, *11* (14), 8848–8871. https://doi.org/10.1021/acscatal.1c01973.

(19) Hu, B.; Kuo, D.-Y.; Paik, H.; Schlom, D. G.; Suntivich, J. Enthalpy and Entropy of Oxygen Electroadsorption on RuO2(110) in Alkaline Media. *J. Chem. Phys.* **2020**, *152* (9), 094704. https://doi.org/10.1063/1.5139049.

(20) Kuo, D.-Y.; Kawasaki, J. K.; Nelson, J. N.; Kloppenburg, J.; Hautier, G.; Shen, K. M.; Schlom, D. G.; Suntivich, J. Influence of Surface Adsorption on the Oxygen Evolution Reaction on IrO$_2$ (110). *J. Am. Chem. Soc.* **2017**, *139* (9), 3473–3479. https://doi.org/10.1021/jacs.6b11932.

(21) Kuo, D.-Y.; Paik, H.; Kloppenburg, J.; Faeth, B.; Shen, K. M.; Schlom, D. G.; Hautier, G.; Suntivich, J. Measurements of Oxygen Electroadsorption Energies and Oxygen Evolution Reaction on RuO$_2$ (110): A Discussion of the Sabatier Principle and Its Role in Electrocatalysis. *J. Am. Chem. Soc.* **2018**, *140* (50), 17597–17605. https://doi.org/10.1021/jacs.8b09657.

(22) Rao, R. R.; Kolb, M. J.; Giordano, L.; Pedersen, A. F.; Katayama, Y.; Hwang, J.; Mehta, A.; You, H.; Lunger, J. R.; Zhou, H.; Halck, N. B.; Vegge, T.; Chorkendorff, I.; Stephens, I. E. L.; Shao-Horn, Y. Operando Identification of Site-Dependent Water Oxidation Activity on Ruthenium Dioxide Single-Crystal Surfaces. *Nat. Catal.* **2020**, *3* (6), 516–525. https://doi.org/10.1038/s41929-020-0457-6.

(23) Kwon, S.; Stoerzinger, K. A.; Rao, R.; Qiao, L.; Goddard, W. A.; Shao-Horn, Y. Facet-Dependent Oxygen Evolution Reaction Activity of $IrO_2$ from Quantum Mechanics and Experiments. *J. Am. Chem. Soc.* **2024**, *146* (17), 11719–11725. https://doi.org/10.1021/jacs.3c14271.

(24) Xu, P.; Von Rueden, A. D.; Schimmenti, R.; Mavrikakis, M.; Suntivich, J. Optical Method for Quantifying the Potential of Zero Charge at the Platinum–Water Electrochemical Interface. *Nat. Mater.* **2023**, *22* (4), 503–510. https://doi.org/10.1038/s41563-023-01474-8.

(25) Bohnen, K.-P.; Heid, R.; De La Peña Seaman, O. *Ab Initio* Lattice Dynamics and Thermodynamics of $RuO_2$ ( 110 ) Surfaces. *Phys. Rev. B* **2010**, *81* (8), 081405. https://doi.org/10.1103/PhysRevB.81.081405.

(26) Feng, T.; Wang, Y.; Herklotz, A.; Chisholm, M. F.; Ward, T. Z.; Snijders, P. C.; Pantelides, S. T. Determination of Rutile Transition Metal Oxide (110) Surface Terminations by Scanning Tunneling Microscopy Contrast Reversal. *Phys. Rev. B* **2021**, *103* (3), 035409. https://doi.org/10.1103/PhysRevB.103.035409.

(27) Karlberg, G. S.; Jaramillo, T. F.; Skúlason, E.; Rossmeisl, J.; Bligaard, T.; Nørskov, J. K. Cyclic Voltammograms for H on Pt(111) and Pt(100) from First Principles. *Phys. Rev. Lett.* **2007**, *99* (12), 126101. https://doi.org/10.1103/PhysRevLett.99.126101.

(28) Hörmann, N. G.; Reuter, K. Thermodynamic Cyclic Voltammograms Based on *Ab Initio* Calculations: Ag(111) in Halide-Containing Solutions. *J. Chem. Theory Comput.* **2021**, *17* (3), 1782–1794. https://doi.org/10.1021/acs.jctc.0c01166.

(29) Rossmeisl, J.; Jensen, K. D.; Petersen, A. S.; Arnarson, L.; Bagger, A.; Escudero-Escribano, M. Realistic Cyclic Voltammograms from *Ab Initio* Simulations in Alkaline and Acidic Electrolytes. *J. Phys. Chem. C* **2020**, *124* (37), 20055–20065. https://doi.org/10.1021/acs.jpcc.0c04367.

(30) Van De Walle, A. Methods for First-Principles Alloy Thermodynamics. *JOM* **2013**, *65* (11), 1523–1532. https://doi.org/10.1007/s11837-013-0764-3.

(31) Van De Walle, A.; Asta, M. High-Throughput Calculations in the Context of Alloy Design. *MRS Bull.* **2019**, *44* (4), 252–256. https://doi.org/10.1557/mrs.2019.71.

(32) Van Der Ven, A.; Bhattacharya, J.; Belak, A. A. Understanding Li Diffusion in Li-Intercalation Compounds. *Acc. Chem. Res.* **2013**, *46* (5), 1216–1225. https://doi.org/10.1021/ar200329r.

(33) Cao, L.; Li, C.; Mueller, T. The Use of Cluster Expansions To Predict the Structures and Properties of Surfaces and Nanostructured Materials. *J. Chem. Inf. Model.* **2018**, *58* (12), 2401–2413. https://doi.org/10.1021/acs.jcim.8b00413.

(34) Mathew, K.; Kolluru, V. S. C.; Mula, S.; Steinmann, S. N.; Hennig, R. G. Implicit Self-Consistent Electrolyte Model in Plane-Wave Density-Functional Theory. *J. Chem. Phys.* **2019**, *151* (23), 234101. https://doi.org/10.1063/1.5132354.

(35) Mathew, K.; Sundararaman, R.; Letchworth-Weaver, K.; Arias, T. A.; Hennig, R. G. Implicit Solvation Model for Density-Functional Study of Nanocrystal Surfaces and Reaction Pathways. *J. Chem. Phys.* **2014**, *140* (8), 084106. https://doi.org/10.1063/1.4865107.

(36) Puchala, B.; Thomas, J. C.; Van der Ven, A. CASM Monte Carlo: Calculations of the Thermodynamic and Kinetic Properties of Complex Multicomponent Crystals. arXiv September 20, 2023. http://arxiv.org/abs/2309.11761 (accessed 2024-09-11).

(37) Nørskov, J. K.; Rossmeisl, J.; Logadottir, A.; Lindqvist, L.; Kitchin, J. R.; Bligaard, T.; Jónsson, H. Origin of the Overpotential for Oxygen Reduction at a Fuel-Cell Cathode. *J. Phys. Chem. B* **2004**, *108* (46), 17886–17892. https://doi.org/10.1021/jp047349j.

(38) *NIST-JANAF Thermochemical Tables*. https://janaf.nist.gov.

(39) Marx, D. Proton Transfer 200 Years after von Grotthuss: Insights from Ab Initio Simulations. *ChemPhysChem* **2006**, *7* (9), 1848–1870. https://doi.org/10.1002/cphc.200600128.

(40) Sato, R.; Ohkuma, S.; Shibuta, Y.; Shimojo, F.; Yamaguchi, S. Proton Migration on Hydrated Surface of Cubic $ZrO_2$: *Ab Initio* Molecular Dynamics Simulation. *J. Phys. Chem. C* **2015**, *119* (52), 28925–28933. https://doi.org/10.1021/acs.jpcc.5b09026.

(41) Suntivich, J.; Hautier, G.; Dabo, I.; Crumlin, E. J.; Kumar, D.; Cuk, T. Probing Intermediate Configurations of Oxygen Evolution Catalysis across the Light Spectrum. *Nat. Energy* **2024**. https://doi.org/10.1038/s41560-024-01583-x.

(42) Rao, R. R.; Kolb, M. J.; Halck, N. B.; Pedersen, A. F.; Mehta, A.; You, H.; Stoerzinger, K. A.; Feng, Z.; Hansen, H. A.; Zhou, H.; Giordano, L.; Rossmeisl, J.; Vegge, T.; Chorkendorff, I.; Stephens, I. E. L.; Shao-Horn, Y. Towards Identifying the Active Sites on $RuO_2$ (110) in Catalyzing Oxygen Evolution. *Energy Environ. Sci.* **2017**, *10* (12), 2626–2637. https://doi.org/10.1039/C7EE02307C.

Supplementary Information

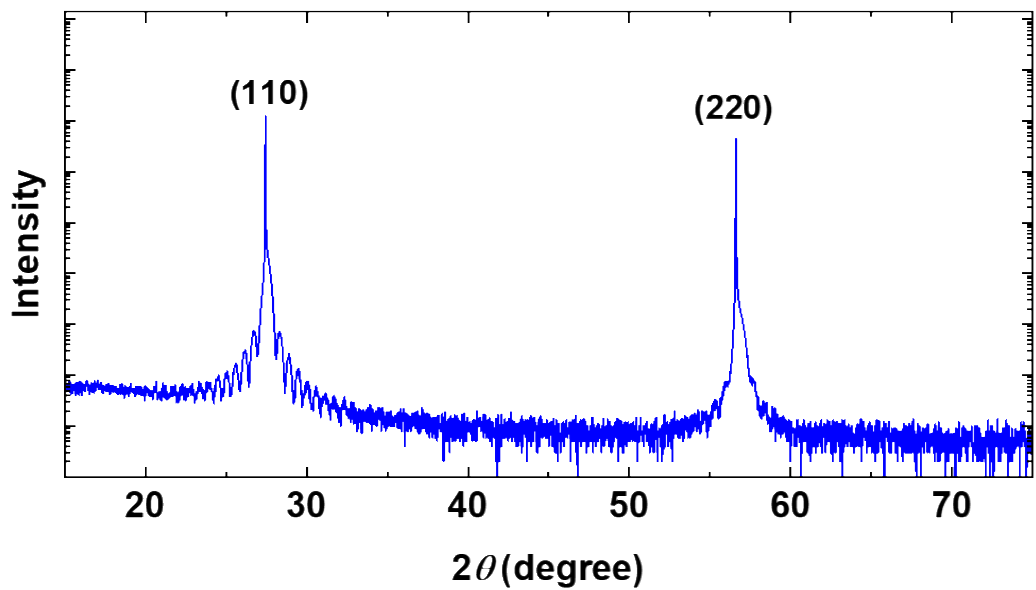

*Figure S 1. X-ray diffraction of an MBE-grown 19 nm RuO$_2$(110) film (~47 formula-units thick) on a TiO$_2$(110) substrate.*

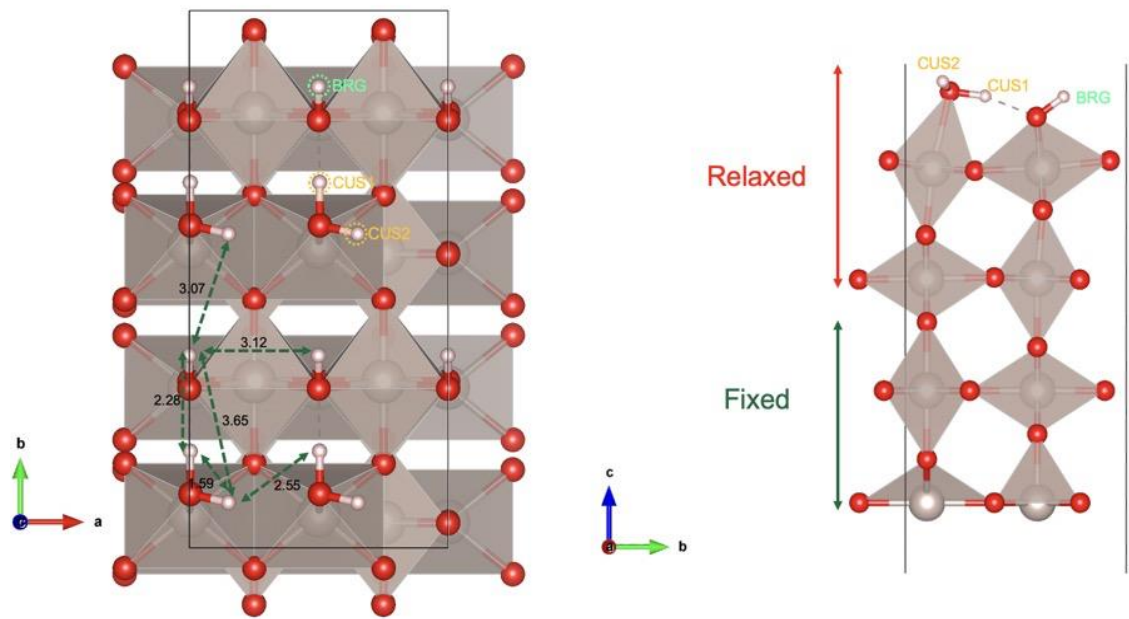

*Figure S 2. (Left) Top view of a 2*2 super-slab of RuO$_2$ (110) surface with fully covered hydrogen atoms (The balls in red, pink are O and H. Oxygen octahedrons are centered with Ru). Two H atoms are on CUS sites with H on CUS Type 1 site pointing towards the BRG site line. Only one H atom can bind on the BRG site with the same pointed direction as CUS1. Some important atomic distances between H atoms are labeled. (Right) Side view of a 1*1 slab of RuO$_2$ (110) with 4-Ru layers. The asymmetric slab with Ru-termination at the bottom is fixed atomic positions as the bulk geometry and O-termination at the top is allowed to relax and adsorbs H atoms.*

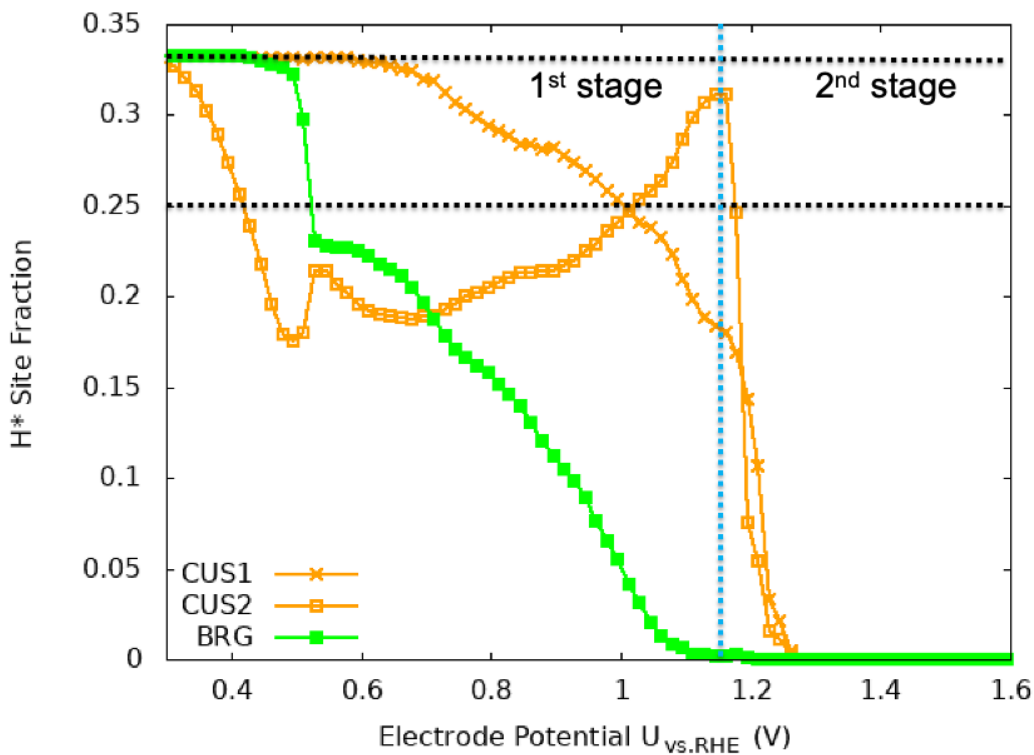

*Figure S 3. *H desorption profile over the three kinds of *H adsorption sites (CUS1, CUS2 and BRG). The two kinds of CUS sites are in orange (with CUS1 in ×, and CUS2 in • ), BRG is in green ■. During the 1$^{st}$ stage, half of CUS1 *H are desorbing, *H on CUS2 desorbs first and fast but get "re-adsorbed" from *H redistributed from CUS1, which maintains a constant H* site-fraction on all CUS; BRG *H are desorbing gradually over the whole course of 1$^{st}$ stage, spanning a potential window of ~1 V and are all gone at ~1.2 V, marking the boundary between 1$^{st}$ and 2$^{nd}$ peak. The 2$^{nd}$ stage starts right after all BRG H* are gone, where a sharp desorption of *H over both CUS sites occurs within a narrow ~0.1 V potential window.*

# Supplementary File

1. Dehydrogenation Movie